\documentclass[manuscript]{acmart}

\setcopyright{rightsretained}
\copyrightyear{2021}
\acmYear{2021}
\acmDOI{}

\acmBooktitle{Co-Designing Resources for Ethics Education in HCI Workshop}

\begin{document}

\title{Teaching Inclusive Engineering Design at a Small Liberal Arts College}

\author{Maggie Delano}
\email{mdelano1@swarthmore.edu}
\orcid{0000-0002-6161-8854}
\affiliation{%
  \institution{Swarthmore College}
  \streetaddress{500 College Avenue}
  \city{Swarthmore}
  \state{Pennsylvania}
  \country{USA}
  \postcode{19081}
}


\begin{abstract}
Modern engineering education tends to focus on mathematics and fundamentals, eschewing critical reflections on technology and the field of engineering. In this paper, I present an elective engineering course and a 3-lecture module in an introductory course that emphasize engaging with the social impacts of technology.
\end{abstract}




\maketitle

\section{Introduction}

Despite an increasing emphasis on ``tech ethics,'' it is not common for engineering students to engage with critical literature related to technology in an engineering class and receive engineering credit for it \cite{fieslerWhatWeTeach2020}. This is due in part to a movement James Malazita dubs the ``Instrumental Turn,'' in which conservative practitioners shifted critiques of technology to separate domains, allowing engineering to be decontextualized \cite{jamesmalazitaEpistemicInfrastructuresInstrumental2018}. This turn also involved a move away from ``shop'' engineering and design toward mathematics and fundamentals. In this paper, I'll discuss how my new course Inclusive Engineering Design seeks to revert some of this decontextualization by explicitly focusing on both critical literature of technology and design projects. I'll also discuss how I've incorporated some of this work into my introductory digital and embedded systems course.

\section{Inclusive Engineering Design}

\subsection{Half Seminar, Half Design Studio}

I describe my Inclusive Engineering Design course as ``half seminar, half design studio.'' Students read critical works in technology studies, including excerpts from \textit{Design Justice} by Sasha Costanza-Chock \cite{costanza-chockDesignJusticeCommunityLed2020}, \textit{The Design of Everyday Things} by Don Norman \cite{norman2013design}, \textit{Do Artifacts Have Politics?} by Langdon Winner \cite{winnerArtifactsHavePolitics1980}, \textit{Algorithms of Oppression} by Safiya Noble \cite{nobleIntroductionPowerAlgorithms2018} and so on.\footnote{For a full list of readings, see the course syllabus \cite{delanoENGR053Fall20202020}.} Each week when reading was assigned, students completed a reflection question related to the reading. In class sessions were ran as workshops or as discussions (``seminar style''). Students also completed two in depth design projects: one related to re-designing a household object, and a second one where the students worked in small groups on a topic of their choosing related to identity. Students also did two small assignments where they attended and reflected on a design event of their choosing, and presented about an example of inclusive design in the media for the class. We had two guest lectures: one about interviewing (Afsaneh Rigot \cite{afsanehrigot2020}), and one about imagining futures (Ashley Jane Lewis \cite{ashleyjanelewis2021}).

\begin{figure}
    \centering
    \includegraphics[width=\textwidth]{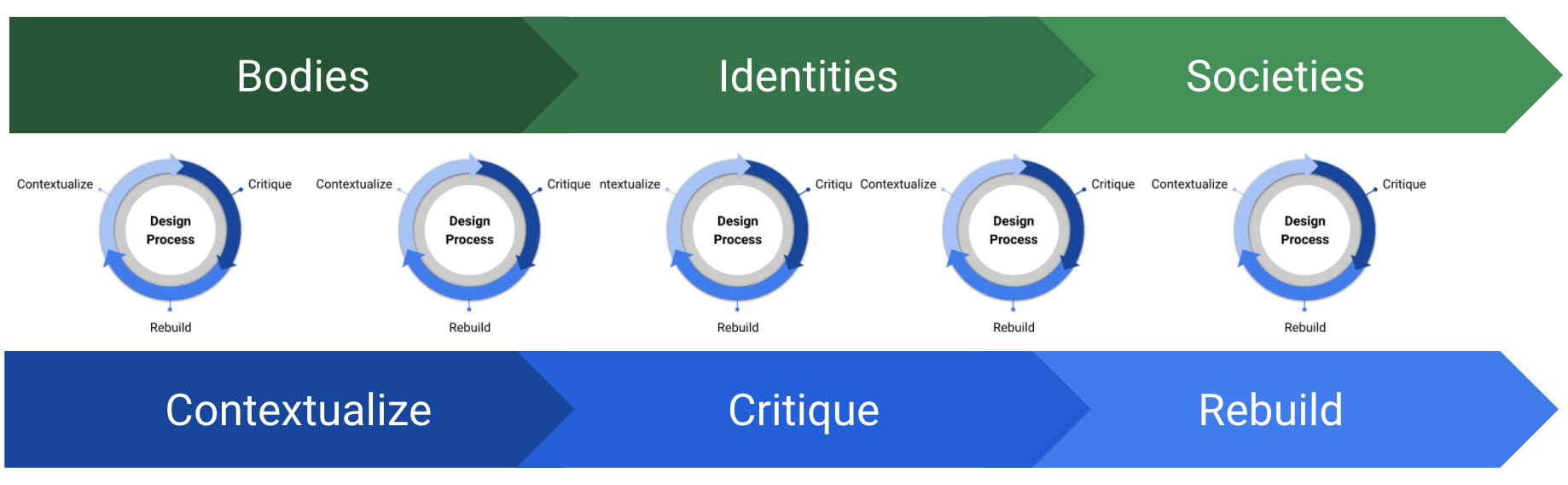}
    \caption{An overview of the units of the course (bodies, identities, society) and the design process associated with it (contextualize, critique, rebuild).}
    \label{fig:designprocess}
\end{figure}

The course was organized into three modules of scale: bodies, identities, societies. Each module included readings and design exercises and projects. Students were also introduced to a design process I came up with called contextualize, critique, rebuild (inspired by a blog post by Sara Hendren \cite{sarahendrenCritiqueRepairCall2020}). These ``sets of threes'' were integrated iteratively over the course of the class (see Figure \ref{fig:designprocess}). I came up with the idea of bodies, identities, societies while reflecting that design assumptions and exclusionary design is easier to conceive of when we think about bodies - that is, it's obvious if a watch or a pair of pants don't fit, or if a wheelchair user cannot navigate a set of stairs without assistance. There might be disagreements on remedies, but not on the fact that certain design assumptions were made and their impact. It's harder to grasp some of the ways these pre-existing biases \cite{friedmanBiasComputerSystems1996} shape the way we think about the technology that we design more broadly - would ankle monitors exist in a society that didn't have prisons \cite{kendraalbertWorldPrison17}? Why do we know more about milking cows than breastfeeding parents \cite{dignazioFeministHCIApproach2016}? By focusing on the body first, we can begin to see how design assumptions manifest, before considering the ways that identities and societies impact and are impacted by design.

A common issue with modern design thinking is the lack of consideration about the context in which one is working \cite{costanza-chockDesignJusticeCommunityLed2020}. Previous implementations, along with what is already working or has been tried in the past, are not always considered. I chose to make evaluating the broader sociotechnical context of an issue at hand to be the first step in my inclusive design process. The second step is critique: to articulate precisely what is not working, and why. Finally, the last step is rebuild. I originally had used the word ``repair'' but I think that ``rebuild'' encourages us to think not just of ``quick fixes'' but to consider how broader structural changes could be facilitated by our designs. 

\subsection{Reception}

A total of 10 students enrolled in the course, 8 of whom were engineering majors/minors and 2 of whom were computer science majors (this is about average for an elective in my department). By the second week, the students were already talking about how important the material we were covering was, and how they were grateful to have a space to discuss the engineering profession and our curriculum specifically. Some students even suggested that this course or one like it become required for all majors. 

Student evaluations at the end of the course were positive. One thread among the more critical feedback was a desire for more advisory input. This was a challenge due to the diversity of projects and because this was the only time the course had been taught. Some of the lack of structure was by design -  I wanted students to try things out and learn for themselves - but some of it was also me lacking expertise in the particular domain or method they were interested in working on. One student suggested more focused ``lecture'' time rather than time for group activities or discussion to help build a more solid foundation in topics that would be applicable to all groups (e.g. how to write interview questions, how to design a survey). Another student suggested that future iterations of the course could focus on outward facing deliverables instead of written project reports (such as a blog post). While I did encourage my students to share their reports with everyone they interviewed and beyond, this is definitely something to consider for future iterations of the course. I would also like to collaborate with other designers and community organizations in future iterations of the course. This is something I wanted to do but decided not to due to the COVID-19 pandemic.

\section{Embedded Systems In Context}

\subsection{Overview}
While my inclusive design course is a dedicated engineering elective, I have also wanted to integrate critical design content into my introductory digital and embedded systems course. The course involves learning to program and how to use basic sensors and actuators with an Adafruit Circuit Playground. I introduced critical design content to this course through a 3 lecture module called ``Embedded Systems in Context'' \cite{delanoL2022EmbeddedSystems2020}. For the first course period, the students work in small groups to research the technical aspects of a topic of their choice. I provide some guiding questions to help them get started (see Appendix \ref{contextAppendix}). The second class period involves researching the social implications of the technology, again with guiding questions. Finally, in the last class of the unit the students give short presentations about their topic to the whole class.

\subsection{Reception}

Reception the first year I offered this module was mixed. I was concerned that students would have difficulty researching in the amount of time available (especially the social impacts aspect), so I decided to pre-assign students to one of four topics with ``seeded'' readings. This meant that students weren't always excited about their topics and the presentations became redundant as multiple groups had the same topics. There was also the perception among the students that the unit had been added as an afterthought to ``fill time'' (despite it existing on the syllabus since the beginning of the semester). This perception was likely due to some pacing issues with the course overall. 

I took the feedback from the students for the second year and allowed students to self-select their own topics. I also emphasized the importance of the assignment and how it was not ``filler.'' While students were working on the assignment there was a campus-wide strike that many students in my class participated in. I spent a class period going into more depth about racial justice and embedded systems to help guide them, and had students turn in videos on a forum instead of presenting for the whole class. Overall, feedback was much more positive.

\section{Conclusion}

It is rare for engineering students to engage with scholarship critical of technology in an engineering class for engineering credit. In this paper, I've presented my experiences working to change that in both a dedicated elective course and in a 3-lecture module in an introductory course. Student reception was positive, so long as the motivation for the work was clear.

\begin{acks}
Thank you to Sara Hendren, Jon Stolk, and Ben Linder, organizers of the Sketch Model Workshop held at Olin College in Summer 2018, where the idea of ``bodies, identities, societies'' was formed, and where I workshopped early ideas for this course. Thank you also to Rose Ridder and Kira Emmons for helping me pilot an earlier version of the Inclusive Engineering Design course as a directed reading in Spring 2020. Thanks to all my students in their respective courses, and thank you to Swarthmore College and the Engineering department for supporting me in this work.
\end{acks}

\bibliographystyle{ACM-Reference-Format}
\bibliography{e53}

\appendix

\section{Inclusive Engineering Design}

\textbf{Official Catalog Course Description:} Technology created by humans reflects our biases and priorities. Engineering a better world requires an interrogation of how we design. This course will combine critical works in technology studies with hands-on, student directed design projects. The course will be divided into three modules that will investigate the relationship between design and bodies, identities, and society. Readings will draw from fields such as disability studies and science and technology studies. Students will apply design methods such as universal design, human centered design, and critical design. This course is open to both Engineering students and non-majors with some previous design experience, such as Computer Science or Art majors.

\textbf{Learning Objectives and Outcomes:}

\begin{enumerate}
    \item In this class, you will:
    \begin{enumerate}
        \item Read contextual and critical works in technology studies
        \item Conduct hands-on, self-directed and supervised design projects
    \end{enumerate}

    \item By the end of this class, you will be able to:
    \begin{enumerate}
        \item Evaluate the impact of a technology (or lack thereof) on individuals, communities, and societies - how are the benefits and harms distributed?
        \item Apply collaborative design methods to address real world problems
    \end{enumerate}
\end{enumerate}
        
\section{Embedded Systems In Context}
\label{contextAppendix}

\textbf{Learning Objectives and Outcomes}

\begin{enumerate}
    \item In this assignment, you will...
    \begin{enumerate}
        \item work in a small team of 2-4 students to research the technical and social aspects of an embedded system of your choice
        \item film a video presentation of you and your teammates presenting the results of your findings to the class
    \end{enumerate}
    \item By the end of this assignment, you should be able to...
    \begin{enumerate}
        \item relate the technical content covered in the course to concrete, real-world examples
        \item discuss how embedded systems shape social and cultural structures, and vice versa
        \item deliver an effective short presentation
    \end{enumerate}
\end{enumerate}

\textbf{Guiding Questions}

\begin{enumerate}
    \item Day 1: Technical Aspects
    \begin{enumerate}
        \item What types of sensors and/or actuators are used by the embedded system?
        \item What types of programming concepts would be helpful in controlling the embedded system?
        \item If you can find examples, what types of microcontroller does the embedded system use?
    \end{enumerate}
    \item Day 2: Social Impacts
    \begin{enumerate}
        \item What does using the embedded system allow a person to do?
        \item Who benefits from using the embedded system, and who is harmed?
        \item Who can use the system and who can't?
    \end{enumerate}
\end{enumerate}

\end{document}